# Mining Fashion Outfit Composition Using An End-to-End Deep Learning Approach on Set Data

Yuncheng Li, LiangLiang Cao, Jiang Zhu, Jiebo Luo, *Fellow, IEEE*

*Abstract*—Composing fashion outfits involves deep understanding of fashion standards while incorporating creativity for choosing multiple fashion items (*e.g.*, Jewelry, Bag, Pants, Dress). In fashion websites, popular or high-quality fashion outfits are usually designed by fashion experts and followed by large audiences. In this paper, we propose a machine learning system to compose fashion outfits automatically. The core of the proposed automatic composition system is to score fashion outfit candidates based on the appearances and meta-data. We propose to leverage outfit popularity on fashion oriented websites to supervise the scoring component. The scoring component is a multi-modal multi-instance deep learning system that evaluates instance aesthetics and set compatibility simultaneously. In order to train and evaluate the proposed composition system, we have collected a large scale fashion outfit dataset with 195K outfits and 368K fashion items from Polyvore. Although the fashion outfit scoring and composition is rather challenging, we have achieved an AUC of 85% for the scoring component, and an accuracy of 77% for a constrained composition task.

## I. INTRODUCTION

Fashion style tells a lot about the subject's interests and personality. With the influence of fashion magazines and fashion industries going online, clothing fashions are attracting more and more attention. According to a recent study by *Trendex North America*[1], the sales of woman's apparel in United States is $111 Billion in 2011 and keeps growing, representing a huge market for garment companies, designers, and e-commerce entities.

Different from well-studied fields including object recognition [1], fashion sense is a much more subtle and sophisticated subject, which requires domain expertise in outfit composition. Here an "outfit" refers to a set of clothes worn together, typically for certain desired styles. To find a good outfit composition, we need not only follow the appropriate dressing codes but also be creative in balancing the contrast in colors and styles. Fig. 1 is an example showing the nontrivial nature of fashion outfit composition. The outfit composition includes a *Felicia Lace mix* dress, a *Nylon* backpack with print, in addition to a pair of *Nicholas Kirkwood Scarp Glitter Mary Janes* shoes. Normally people do not pair a fancy dress with a casual backpack, however, once the shoes were in the outfit, it completes the look of a nice and trendy outfit.

[1]http://goo.gl/QdrAvQ

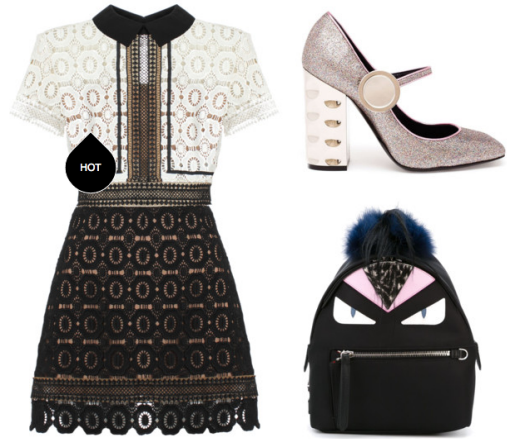

Fig. 1: An example showing the challenging of fashion outfit composition. Normally one would not pair a fancy dress (as in the left) with the casual backpack (as in the bottom right) but once you add in the shoes (as in the top right), it completes the look of a nice outfit.

Although there have been a number of research studies [2] [3] [4] on clothes retrieval and recommendation, none of them considers the problem of fashion outfit composition. This is partially due to the difficulties of modeling outfit composition: On one hand, a fashion concept is often subtle and subjective, and it is nontrivial to get consensus from ordinary labelers if they are not fashion experts. On the other hand, there may be a large number of attributes for describing fashion, for which it is very difficult to obtain exhaustive labels for training. As a result, most of the existing studies are limited to the simple scenario of retrieving similar clothes, or choosing individual clothes for a given event.

This paper proposes a data-driven approach to train a model that can automatically compose suitable fashion outfit. This approach is motivated by the recent surge of online fashion communities, including Polyvore, Pinterest, and YouTube videos, which have greatly helped spreading fashion trends and fashion tips, creating an online culture of sharing one's style with other Internet and mobile users. Such online communities can be very big. For example, Polyvore received 20 million unique monthly visitors in May 2014. By actively interacting with the websites, the users express their opinions on which fashion outfits are good and which are not so well composed. By aggregating the wisdom of the crowds, we obtain user engagement scores (popularity), for the fashion outfits, which are used to train a classifier to score new fashion outfit candidates. The full automatic composition system is built upon the scorer by iteratively evaluating all possible outfit candidates.

There are two main challenges in building this outfit scorer:

1) Complicated visual contents of a fashion image. There are potentially many attributes corresponding to different visual aspects: categories, colors, coherence, patterns, general pairing codes, creative pairing choices, as well as styles for different ages and individuals. It is impossible to label or even list all possible attributes for every clothing image.

2) Rich contexts of fashion outfit. Clothing outfits can reflect personality and interests. Clothing may reflect the style and tradition of different brands and designers, or even the culture of a specific group. To infer such information, we must take into account not only the pixel information but also the context information in the fashion outfit.

To solve the first challenge, we propose an end-to-end system of encoding visual features using a deep convolutional network, which can take a fashion outfit as the input and predict the user engagement levels. To address the second challenge, we propose a multi-modal deep learning framework, which exploits the context information from image, title and category. As the experiments show, the multi-modal approach significantly outperforms a single modality, and provides a more reliable solution for the fashion outfit scoring task, and thus, the full composition tasks.

In summary, our contributions are three folds:

1) We present a novel problem: fashion outfit composition. The manual outfit composition is a very challenging intellectual process, and we hope our system can be of help to fashion designers and fans.

2) We propose a fashion outfit composition solution based on a reliable fashion outfit quality predictor. Predicting fashion outfit quality is an extremely challenging problem, because many interleaving factors, visible or hidden, contribute to the process. The combinatorial nature of the problem also makes it very interesting to serve as a test stone for the state of the art machine learning systems.

3) We propose an end-to-end trainable system to fuse signals from multi-level hybrid modalities, including the images and meta-data of the fashion items, and the information across the fashion items.

4) We collect a large-scale dataset for the fashion outfit related research. This dataset contains fashion items and outfits, associated rich context information, and will be released for future research by the community.

## II. Related work

Fashion domain is a very important and lucrative application of computer vision [5], [6]. The majority of research in this domain focus on fashion image retrieval [7], [8], [9], and fashion image attribute learning [4], [10], [11], [12]. There are also studies on evaluating the compatibility between fashion items [3], [2]. Specifically, Veit *et al.* proposed to learn clothing matches from the Amazon co-purchase dataset [3], and Iwata *et al.* proposed a topic model to recommend *"Tops"* for *"Bottoms"* [2]. Comparing with the previous works on fashion image retrieval or clothing style modeling, the goal of this work is to compose fashion outfit automatically, which has its own challenges in modeling many aspects of the fashion outfits, such as compatibility and aesthetics.

The techniques developed in this paper belong to the category of set classification. Unlike our work, which focuses on fashion related sets, most of the existing work was applications of face recognition from video frame sequences and multi-view object recognition. Recently, Huang *et al.* proposed a parametric manifold distance based on tangent map, and used metric learning to learn a problem adaptive manifold metric to perform kNN classification [13]. Lu *et al.* proposed a novel loss function to train Convolutional Neural Networks for the image set classification problem [14]. Li *et al.* combined dictionary learning and metric learning to recommend online image boards [15]. The key differences of this work from previous image set classification are the following: 1) We are the first to consider fashion outfits. The goal of the fashion outfit modeling is to disentangle the style factor hidden among the fashion items, while the previous works [13], [14] focuses on the cases where the images share the same object class. 2) Along with the images, we integrate the meta-data of the fashion items to further improve the fashion outfit modeling, for which we propose to jointly learn modality embedding and fuse modalities.

Our work is partially motivated by the previous work in multimedia understanding [16], [17], [18], [19], [20], [21], [22]. Most of these works suggest leveraging visual analysis with other modalities such as text and audio information. Moreover, the recent progress of deep neural networks in visual recognition [23], [24], [25] and natural language processing [26], [27], [28] has shown that the recognition performance in both fields have been greatly improved. Inspired by these progresses, this paper tries to validate that we can achieve better results by combining the-state-of-the-art techniques in both fields.

## III. Methodology

In this section, we first present our algorithm for fashion outfit composition based on an outfit scoring model, and then we present our multi-modal multi-instance scoring model.

### A. Fashion Outfit Composition

Let $\mathbf{I}$ denotes the set of all fashion items, $\mathbf{S}_i$ denote an outfit, and $x_{i,j} \in \mathbf{I}$ denote the items in the outfit $\mathbf{S}_i$, so that $\mathbf{S}_i = \{x_{i1}, x_{i2}, \ldots, x_{i|\mathbf{S}_i|}\}$. Each item $x_{i,j}$ is associated with multiple modalities, such as image, title and category.

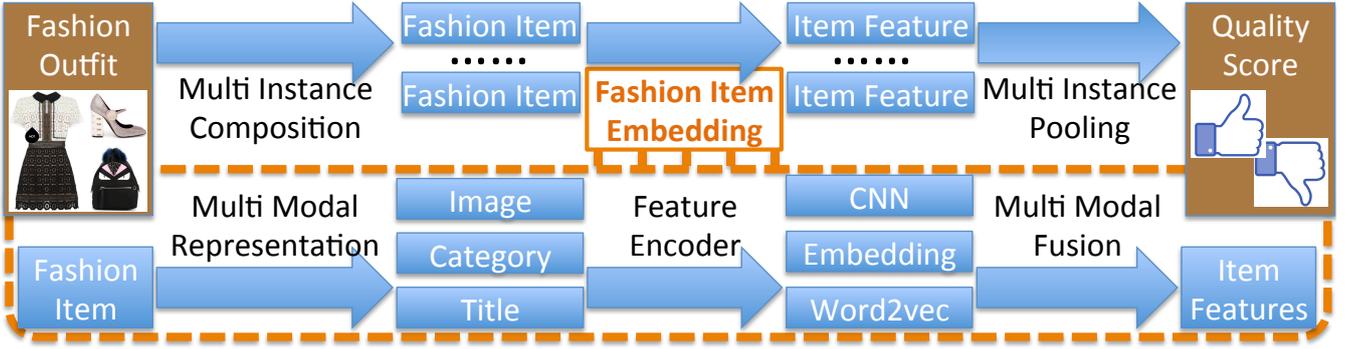

Fig. 2: The proposed fashion outfit scoring model.

---

**Algorithm 1:** Fashion outfit composition

**Input** : Outfit scoring model $f(\mathbf{S}_i; \theta)$
**Input** : Candidate fashion items $\mathbf{I}$
**Input** : Seed fashion outfit $\mathbf{S}^0$
**Input** : Target outfit length $L$
**Output** : New fashion outfits $\mathbf{S}^*$

1 $\mathbf{S}^* \leftarrow \mathbf{S}^0$
2 **repeat**
3 $\quad$ $t^* \leftarrow \arg\max_{t \in \mathbf{I}} f(\{t\} \cup \mathbf{S}^*)$
4 $\quad$ $\mathbf{S}^* \leftarrow \{t^*\} \cup \mathbf{S}^*$
5 **until** $|\mathbf{S}^*| = L$;

| Notation | Meaning |
|---|---|
| $\mathbf{S}_i$ | a fashion outfit |
| $y_i$ | quality label for $\mathbf{S}_i$ |
| $\mathbf{I}$ | the set of all fashion items |
| $x_{i,j} \in \mathbf{I}$ | $j^{th}$ item in the outfit $\mathbf{S}_i$ |
| $x_{i,j}^k \in x_{i,j}$ | $k^{th}$ modality of the item $x_{i,j}$ |
| $f(\mathbf{S}_i)$ | quality score for outfit $\mathbf{S}_i$ |
| $E^k(x_{i,j}^k)$ | feature encoder for the $k^{th}$ modality |
| $F(x_{i,j})$ | feature fusion model for the fashion item $x_{i,j}$ |
| $P(\mathbf{S}_i)$ | feature pooling model for the fusion outfit $\mathbf{S}_i$ |
| $C(\mathbf{S}_i)$ | outfit quality classification model |

TABLE I: Definition of the notations

Let $x_{i,j}^k$ denote the $k^{th}$ modality of the item $x_{i,j}$, so that $x_{i,j} = \{x_{i,j}^1, x_{i,j}^2, \ldots, x_{i,j}^K\}$, where $K$ is the total number of modalities.

The problem of fashion outfit composition is formulated as an iterative item selection process, presented in the Algorithm 1. In the algorithm, the scoring model $f(\mathbf{S}_i) \in R$ indicates how well the outfit $\mathbf{S}_i$ is composed, and we will discuss its design in the next section. The set $\mathbf{I}$ denotes the candidate fashion items in the database. The seed outfit $\mathbf{S}^0$ indicates the items that have already been chosen by the users. The target outfit length $L$ denotes how many items should be in the final outfit.

For example, a user may want to find the best *"Skirt"* that goes well with his/her current *"Shoes"*, *"Hat"* and *"Top"*. In this example, all *"Skirt"* in the database is the candidate item set $\mathbf{I}$, the current *"Shoes"*, *"Hat"* and *"Top"* are the seed outfit $\mathbf{S}^0$, and the length of the target outfit is 4. The Algorithm 1 allows flexible applications, and the core component is the scoring function $f(\mathbf{S}_i; \theta)$, which we discuss in the next section.

### B. Fashion Outfit Scoring Model

In this section, we discuss the architecture of the scoring model $f(\mathbf{S}_i; \theta)$, and the supervised information used to learn the model parameters. The scoring model consists of:

1) Feature encoders for each modality $E^k : x_{i,j}^k \mapsto R^d$, where $d$ is the feature dimension;
2) Fusion model to integrate the multiple modalities $F : x_{i,j} \triangleq \{E^k(x_{i,j}^k), k = 1 \ldots K\} \mapsto R^d$, where $K$ is the total number of modalities (in this paper, $K = 3$);

3) Pooling model to map the outfit into a single feature vector $P : \mathbf{S}_i \triangleq \{x_{i,j}, j = 1 \ldots |\mathbf{S}_i|\} \mapsto R^d$;
4) Classification model to perform prediction $C : P(\mathbf{S}_i) \mapsto \hat{Y}_i \in \{0, 1\}$, where 1 denotes that $\mathbf{S}_i$ is a good outfit. The classifier output is used as the $f(\mathbf{S}_i)$ in the composition algorithm for new fashion outfits.

Table I list all the notations for easy reference. Fig. 2 gives an overview of the fashion outfit scoring model. These components are connected and trainable from end to end, and the training outfit is a list of pairs $(\mathbf{S}_i, y_i)$, where $y_i$ is the label telling if $\mathbf{S}_i$ is well composed ($y_i = 1$) or not ($y_i = 0$). In the following, we explain each component separately.

***Visual Feature Encoding***

As shown in the fashion item example (Fig. 1), the fashion item image contains the most important information needed for outfit composition, and we encode the item image $x_{i,j}^1$ with Convolutional Neural Networks (*ConvNets*). There are many ConvNets architectures to choose from, and we use a variant of AlexNet for simplicity. The AlexNet variant, is the same as the original, except that we do not use pretrained weights and we replace local response normalization with batch normalization. We use the output of the *fc6* layer as the encoding feature of the input image. The dimension of the image encoding is 4096.

***Context Feature Encoding***

While images contains the most important information for outfit composition, the state of the art image understanding fails to capture all the details in the images for composition inference. We employ the context information, including the item title and category as a remedy to help fashion item modeling.

*GloVe* model is used to encode the item title $x_{i,j}^2$ [27]. *GloVe* model is a distributed word representation that maps words to a 300d vector space. The vector space captures the word semantics, and has been used in a variety of text analysis applications. Given an item title,

1) A basic tokenizer is used to chop it into tokens;
2) Tokens are mapped to 300d vectors using the *GloVe* model;
3) Average the vectors into a single 300d vector.

While there are more sophisticated methods using Recurrent Neural Networks (RNN) or Convolutional Neural Networks (CNN) to combine the GloVe vectors [29], we employ simple average pooling to aggregate the GloVe vectors for simplicity.

*Categorical embedding* is used to encode the item category $x_{i,j}^3 \in \{1 \ldots CC\}$, where $CC$ is the total number of categories. Basically, *Categorical embedding* is a lookup table that maps the item category (represented by integers) into a trainable vector. The dimension of the category embedding is 256.

### Multi-modality Fusion Model

Multiple modalities contain complementary information for the final task. In the following, we explain our pipeline to fuse the multiple modalities:

$$F : \{E^1(x_{i,j}^1), E^2(x_{i,j}^2), \ldots, E^3(x_{i,j})\} \mapsto F_{i,j} \in R^d \quad (1)$$

The fusion model $F_{i,j}$ is designed as follows,

1) Using a single layer perceptron, reduce the dimension of the feature vector $E^k(x_{i,j}^k) \in R^{d_k}$ to the same size $d$: $E^k(x_{i,j}^k) \mapsto \tilde{E}^k(x_{i,j}^k) \in R^d$;
2) Concatenate the vectors $\{\tilde{E}^k\}$ into a single vector $R^{Kd}$, where $K$ is the total number of modalities;
3) Apply 2-layer MLP (multiple layer perceptron) with ReLU nonlinearity and dropout to obtain the final item representation $F_{i,j} \in R^d$.

### Multi-instance Pooling Model

Using the feature encoders and the fusion model, the fashion items are embedded into the $F_{i,j} \in R^d$. A fashion outfit is composed of multiple fashion items, i.e., $\mathbf{S}_i \triangleq \{x_{i,j}, j = 1 \ldots |\mathbf{S}_i|\}$, and the pooling model $P$ maps the outfit into a single vector $P_i \in R^d$:

$$P : S_i \triangleq \{x_{i1}, x_{i2}, \ldots x_{i|S_i|}\} \mapsto P_i \in R^d \quad (2)$$

In practice, the pooling function should handle variable length $\mathbf{S}_i$, and respect the fact that the input is an order-less set. In the experiments, we find that non-parametric element wise reduction works very well, while satisfying the requirements and being very efficient. The non-parametric reduction $r : \{p \in R\} \mapsto R$ is based on commutative functions, such as *add*, *max* or *prod*. For example, the mean reduction $r^{mean}$ takes an outfit of numbers and outputs the *mean* value at each dimension, the maximal reduction $r^{max}$ takes an outfit of numbers and output the *maximal* value. Applying the reduction function $r$ for each dimension independently, the vector outfit $\{R^d\}$ is mapped to a single vector $R^d$.

The pooling model $P$ aggregates information from individual fashion items to produce a holistic embedding for the fashion outfit. Empirically, we observed that the element wise reduction function works very well. However, there are other more advanced models that work very well to deal with variable length inputs, such as Recurrent Neural Networks (RNN) [30]. In the following, we explain the RNN based pooling models.

| | train | dev | test | total |
|---|---|---|---|---|
| #outfits | 156,384 | 19,407 | 19,571 | 195,262 |
| #items | 294,782 | 36,677 | 36,792 | 368,251 |

TABLE II: Basic statistics for the dataset splits.

Recurrent Neural Networks (RNNs) achieved the state of the art performance on image/video captioning, language modeling, and machine translation, by modeling the dependency within time series. We adapt RNN as a pooling model to encode the variable length fashion items. Specifically, given a fashion outfit $\mathbf{S}_i \triangleq \{x_{i,j}, j = 1 \cdots |\mathbf{S}_i|\}$, RNN maintains a state $h_{i,j} \in R^d$, and performs the following state update for each fashion item:

$$h_{i,j} = tanh(h_{i,j-1}W_h + x_{i,j}W_x + B), \quad (3)$$

where initial state $h_{i,0}$ is fixed as all zero vector, $x_{i,j}$ is the embedding of the fashion item (given by the feature encoders and fusion models), $W_h$, $W_x$ and $B$ are model parameters, and the final state $h_{i|\mathbf{S}_i|}$ is used as the embedding of the fashion outfit $\mathbf{S}_i$.

### Outfit Classification Model

Given the feature encoders, fusion model and pooling model, the fashion outfit $\mathbf{S}_i$ is mapped to a fixed length vector $s_i \in R^d$, which is used in the following classification model to evaluate the quality of the outfits.

$$Pr(\hat{Y}_i = 1|\mathbf{S}_i) = \delta(w^T s_i + b), \quad (4)$$

where the $w^T$ and $b$ are the weights and bias of the linear model, respectively, and $\delta(s)$ is the sigmoid function $1/(1 + exp(-s))$.

To learn the parameters in the framework, a loss function $l_i$, based on cross entropy, is defined on the training data $(\mathbf{S}_i, y_i)$,

$$l_i = y_i Pr(\hat{Y}_i) + (1 - y_i)(1 - Pr(\hat{Y}_i)), \quad (5)$$

where the label $y_i$ is the ground truth outfit quality measurement to supervise the entire network. As will be detailed in the Experiment section, the ground truth label $y_i$ is defined as the user engagement level of the outfit $S_i$ on a popular fashion oriented website.

As shown in the Fig. 2, all these models are connected together and the gradients can be backpropagated to learn the model parameters, so we end up with a trainable system to learn all the components adaptively. The learnable parameters in the system includes the convolution and linear layer weights in the image encoder $E^1(x_{i,j})$, the categorical embedding vectors in the item category encoder $E^3(x_{i,j})$, the weights of the MLP layers in the fusion model $F$, the parameters in the pooling model $P$ and the linear classification model $C$.

## IV. Experiments

In this section, we discuss the data collection, the evaluation of the scoring model and the composition model, and some further analysis.

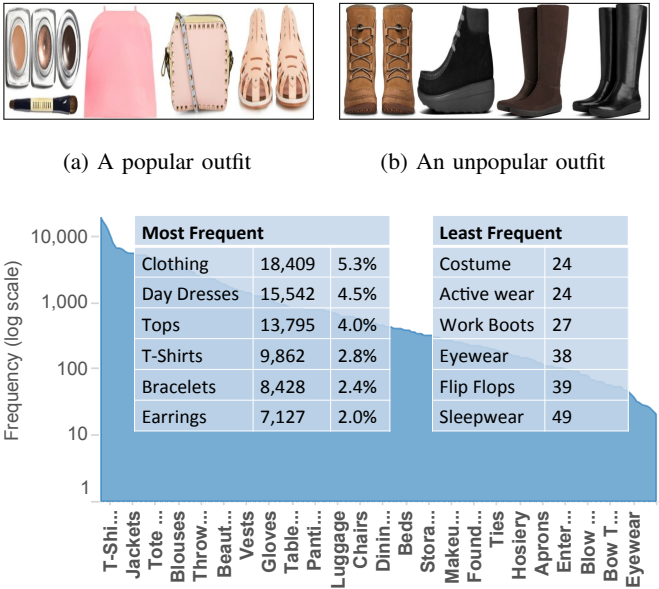

(a) A popular outfit      (b) An unpopular outfit

| Most Frequent | | | Least Frequent | |
|---|---|---|---|---|
| Clothing | 18,409 | 5.3% | Costume | 24 |
| Day Dresses | 15,542 | 4.5% | Active wear | 24 |
| Tops | 13,795 | 4.0% | Work Boots | 27 |
| T-Shirts | 9,862 | 2.8% | Eyewear | 38 |
| Bracelets | 8,428 | 2.4% | Flip Flops | 39 |
| Earrings | 7,127 | 2.0% | Sleepwear | 49 |

Fig. 4: Category frequency and examples. The X-axis iterates the categories. The overlaid tables show more category examples and their actual frequency in our dataset, and the third column of the table on the right lists the percentage frequency.

### A. Dataset

We collect a dataset from Polyvore.com, the most popular fashion oriented website based in US. Polyvore provides tools, templates and friendly interfaces for users to create fashion outfits, and tens of thousands of fashion outfits are created everyday. Fig. 1 shows an example of user curated fashion outfits. As shown in the example, the fashion items are carefully and beautifully organized to demonstrate specific fashion styles. These fashion outfits are viewed, favored and recreated by visitors, and some of the fashion outfits attract high volume of attention. Fig. 3a and 3b are fashion outfit examples. From Polyvore.com, we crawl fashion outfits, which are associated with the number of likes and a number of fashion items. For each fashion item, we crawl the image, the title, and the category. Take the outfit in Fig. 1 as an example, the title for the dress, the pumps and the bag are "Self-Portrait Felicia Lace Mix Dress", "Nicholas Kirkwood Scarpa Glitter Mary Jane Shoes", and "Fendi Nylon Back Pack With Print", respectively. The categories for the three items are "Cocktail Dresses", "Shoes", and "Bags", respectively.

We perform some simple filtering over the crawled datasets to clean up the raw dataset.

1) Each fashion item is associated with a category, such as *"Flip Flops"* and *"Men's Bracelets"*. We remove the categories associated with fewer than 500 items.
2) Remove items that appear in more than 5 outfits. This is to avoid the model overfitting to a particular item.
3) Given the like count for the fashion outfits, we obtain the $1^{th}$, $40^{th}$, $90^{th}$ and $99^{th}$ percentiles. We label the outfits with a like count between $1^{th}$ and $40^{th}$ as unpopular, and those with a like count between $90^{th}$ and $99^{th}$ as popular outfits. Therefore, we have four times more unpopular outfits than popular outfits. We throw away outfits that do not fall into either range, because they are either outliers or uncertain.
4) We segment the entire fashion outfits into training, development, and testing splits. For this segmentation, we make sure there is no overlap between splits, so that the items in the testing split are never seen in the training. To achieve this, we construct a graph with the fashion outfits as nodes, and if two fashion outfits have a common item, we draw an edge between the corresponding nodes. After graph segmentation based on connected components, we obtain the fair training/development/testing splits.
5) To simplify the data pipeline and without loss of generality, we fix the number of items in the fashion outfit as 4. For outfits with more than four items, we randomly sample up to 5 random combinations and treat them as independent 4-item fashion outfits.

After the filtering process, the number of fashion outfits and the fashion items are listed in Table II for each train/development/test splits. After the filtering process, the total number of item categories is 331. The most frequent category has 18,407 items, while the least frequent category has only 21 items. The category examples and the number of items for each category are shown in Fig. 4.

### B. Evaluate Fashion Outfit Quality Scorer

The quality scorer $f(\mathbf{S}_i)$ determines the composition quality in the Algorithm 1. In this section, we present the performance of the scorer model and the detailed ablation analysis on the key model parameters.

#### Evaluation Protocols

The train split is used to train the model parameters for the scorer $f(S_i)$, the development split is used to select hyper-parameters, such as the embedding dimension $d$ for the scorer. The testing split is used as ground truth for evaluation. Because the scorer is essentially a classification model, we use two classification metric for evaluation: Area Under Curve (AUC) of the ROC and Average Precision. There are a few hyper-parameters in the proposed models, such as the modality combination, the embedding dimension $d$, the pooling model $P$ and the number of iterations. In the following sections, unless specified otherwise, we analyze each of them separately, and when analyzing a specific parameter, we use the development split to select the other parameters and report the performance on the testing split.

#### Comparison with existing methods

There are two directions weakly related to this work: image set classification based on metric learning [13], and clothing style learning using Siamese network [3]. In the following, we explain how we adapt these methods as fashion outfit scoring model, and make comparison with our proposed method.

The **LEML** method, proposed in [13], is a state of the art algorithm on image set classification. The traditional image set classification methods assume that the images in a set reside in a structured manifold, and propose various manifold metrics based on the image feature co-variance. Specifically, LEML is a metric learning method to learn distances between image set pairs $\mathbf{S}_i$ and $\mathbf{S}_j$. Given tuples of $(\mathbf{S}_i, \mathbf{S}_j, y_{i,j})$, where $y_{i,j} \in \{0, 1\}$ tells whether $\mathbf{S}_i$ and $\mathbf{S}_j$ are from the same class ($y_{i,j} = 1$ if yes), LEML learns a distance metric $d(\mathbf{S}_i, \mathbf{S}_j)$, such that $d(\mathbf{S}_i, \mathbf{S}_j)$ is small for $y_{i,j} = 1$, and $d(\mathbf{S}_i, \mathbf{S}_j)$ is large for $y_{i,j} = 0$. We refer to the original paper [13] for more details, such as the parametric forms of the $d(\mathbf{S}_i, \mathbf{S}_j)$ and the optimization objectives. The complexity of the original LEML algorithm is in the order of the number of training samples, so we adapt it with stochastic approximation to make it work with the large scale problem presented in this paper. In addition, LEML algorithm works on static features, so we need to extract visual features from the images. For this purpose, we extract the widely used semantic features using the AlexNet network [31] pretrained on the ImageNet object recognition task [1]. Given a new fashion outfit, the nearest neighbor training samples are retrieved in the learned metric space, and the labels of the neighbors are voted to obtain the quality label for the new fashion outfit.

|  | LEML [13] | Siamese [3] | Ours (image only) |
|---|---|---|---|
| AUC | 0.616 | 0.601 | **0.757** |
| AP | 0.237 | 0.169 | **0.364** |

TABLE III: Comparison with existing methods.

There are existing systems on using end-to-end trainable ConvNets to match clothing images [3] using the Siamese architecture and contrastive loss function. Siamese architecture can also be used as the scoring model. Given a training sample $(\mathbf{S}_i, y_i)$, we randomly select one item $x_{i,j'}$ as the pivot item, and the rest items $\mathbf{Q}_i \triangleq \mathbf{S}_i - \{x_{i,j'}\}$ as the query set. The pivot item $x_{i,j'}$ and the query set $\mathbf{Q}_i$ should be a good match if the outfit $S_i$ is of high quality. Therefore, we can build the tuple $(\mathbf{Q}_i, x_{i,j'}, y_i)$ to train the Siamese network, using the contrastive loss function:

$$l(\theta; \mathbf{Q}_i, x_{i,j'}, y_i) = y_i d_i^2 + (1 - y_i)max(\alpha - d_i, 0)^2$$
$$d_i \triangleq d(\theta; \mathbf{Q}_i, x_{i,j'}) = \|\mathbf{Q}_i(\theta) - x_{i,j'}(\theta)\|, \quad (6)$$

where $\theta$ are the model parameters, $\alpha$ is the margin to control sensitivity (fixed at 10 in the experiments). The item encoders and pooling models, as explained in the Section IV-B, are used represent $x_{i,j'}$ and $\mathbf{Q}_i$, respectively, so the Siamese model contains the same set of model parameters as the proposed outfit scorer $f(\mathbf{S}_i)$.

Given a new fashion outfit $\mathbf{S}_i$, we build the pair $(\mathbf{Q}_i, x_{i,j'})$ by picking an arbitrary item as the pivot item $x_{i,j'}$ and the remaining items as the query $\mathbf{Q}_i$. Then, we use the negative distance $-d(\mathbf{Q}_i, x_i; \theta)$ as the quality score of the outfit $\mathbf{S}_i$. For fair comparison, we use the same number of iterations, optimization method and tuning for both the proposed framework and the *Siamese* baseline.

In Table III, we compare the proposed models with the LEML model and Siamese architecture. To make a fair comparison, we use only images in this experiment. From this result, we make the following observations:

1) Compared with LEML, which is a state of the art image set classification method based on manifold metric learning, our proposed framework benefits from end-to-end representation learning. Once a good representation is learned, our methods

|  | AUC | AP |
|---|---|---|
| image | 0.757 | 0.364 |
| category | 0.778 | 0.343 |
| image+category | 0.818 | 0.414 |
| title | 0.820 | 0.455 |
| image+title | 0.822 | 0.476 |
| title+category | 0.835 | 0.476 |
| image+title+category | **0.852** | **0.488** |

TABLE IV: Comparison of different modality combinations.

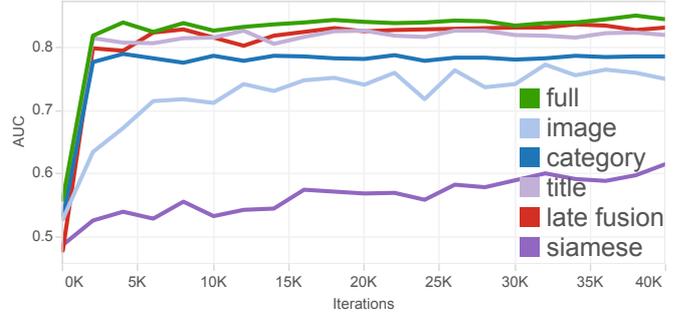

Fig. 5: Convergence property with different modalities. (Best viewed in color)

employ simple strategy to aggregate the evidence from the fashion items.

2) Compared with the Siamese network, our proposed framework solves the outfit scoring problem directly by classification, and thus outperforms the Siamese network. The Siamese underperforms partly because the matching objective is not perfectly aligned with the outfit quality objective.

### Compare Different Modalities

The item image, title and category are available for the fashion outfit scoring model. In this section, we compare the strength of each modality and their combinations to predict fashion outfit quality level. Given a total of three modalities, there are seven modality configurations. The performance is compared in Table IV, which shows that,

1) Fashion item title works best on its own, but image and category add complementary information.

2) Category alone does not work very well, partly because many of the items are labeled as very high level categories. Specifically, Fig. 4 shows statistics of the item categories, and the top frequent categories are *"Clothing"*, *"Day Dresses"*, *"Tops"* and *"T-Shirts"*.

3) The ConvNets used in this paper contains millions of parameters, but the number of train samples is relatively small compare the ImageNet challenges, which partly explains that the image only models performs badly.

Fig. 5 shows the convergence property with different modality combinations. The models converge well, except for the *Siamese* baseline, which is commonly known to converge slowly [3].

### Compare Different Pooling Model $P$

The pooling model $P$ is used to aggregate information from all the items, and potentially model the interaction between the items. We compare three different pooling models, *mean reduction*, *max reduction*, and *RNN reduction*, which are explained in the Section IV-B. The performance is compared in Table V, which shows that the mean reduction achieved the best result, but RNN performs poorly, partly because it enforces a specific order over the fashion items, but the fashion outfit is orderless.

| | mean | max | RNN |
|---|---|---|---|
| AUC | **0.852** | 0.839 | 0.813 |
| AP | **0.488** | 0.476 | 0.416 |

TABLE V: Comparison of different pooling models.

| | max | min | mean | ours |
|---|---|---|---|---|
| AUC | 0.797 | 0.812 | 0.836 | **0.852** |
| AP | 0.374 | 0.406 | 0.444 | **0.488** |

TABLE VI: Comparison with late fusion models

| | 1 | 2 | 4 | 8 | 16 | 32 | 64 | 128 |
|---|---|---|---|---|---|---|---|---|
| AUC | 0.813 | 0.834 | 0.840 | **0.852** | 0.841 | 0.841 | 0.842 | 0.848 |
| AP | 0.43 | 0.452 | 0.452 | **0.488** | 0.487 | 0.484 | 0.465 | 0.467 |

TABLE VII: Performance with different embedding dimensions $d$

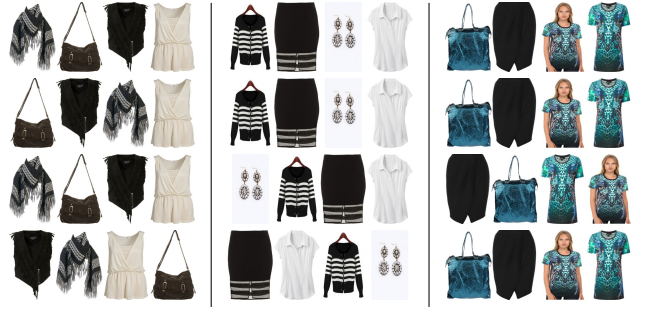

TABLE VIII: Fashion item importance analysis. Each column is one fashion outfit, and the items are sorted according to the item importance (explained in Sec. IV-D). Important items are listed first. The four rows corresponds to the full, image, title, categories modality configurations, respectively.

*Compare with Late Fusion*

In this experiment, we compare the proposed method in Section IV-B with a late fusion baseline, in order to highlight the importance to consider items in a outfit collectively at the training time. Specifically, we transform the original training data $(\mathbf{S}_i, y_i)$ into item level labeled data $\{(x_{i,j}, y_i)|x_{i,j} \in S_i\}$, and use the item level labeled data to train item level quality classifier. In the testing time, the quality scores of individual items are aggregated using mean/max/min to obtain the set level quality score. The late fusion model completely ignores the interaction among the items. The results are compared in the Table VI, which shows that the late fusion based on mean reduction works best, but still underperforms than our proposed model. This experiment result shows that it is important to model the interaction between the items in the training time.

*Compare Different Embedding Dimension* $d$

Using the feature encoders and fusion model, the fashion items are embedded into a fixed length vector of dimension $d$. The performance with different embedding dimensions are listed in Table VII, which shows that the performance peaks at $d = 8$. Too small or large embedding size will hurt the performance.

### C. Evaluate Outfit Composition Algorithm

Once the fashion outfit scorer is trained, the outfit composition algorithm (Algorithm 1) can be used to compose fashion outfits automatically. As with the evaluation of the scorer, the evaluation of the composition model is also based on the test splits explained in Section IV-A. In the following, we first explain the evaluation protocol, and then discuss the results.

*Constrained Composition Setup*

By design, the Algorithm 1 is too flexible such that we can only perform partial evaluation. We propose to following constrained setup that is feasible to evaluate,

1) The outfit scoring model $f(\mathbf{S}_i)$ is trained and fixed as the framework explained in Section IV-B;
2) The candidate set $\mathbf{I}$ contains five randomly sampled items;
3) Seed fashion outfits $\mathbf{S}^0$ are three item outfits.
4) Target outfit length $L$ is fixed as 4;

In other words, the constrained setup simulate the use case that the user wants to find the best item that goes well with the current three item outfit $\mathbf{S}^0$. The target outfit length $L$ is fixed at 4, because the outfit in the test split is with length 4, *i.e.* $|\mathbf{S}_i| = 4$.

Because of the constraints, the arguments of the Algorithm 1 reduced from four to two, and the functional signature of the constrained composition algorithm can be written as:

$$
\begin{aligned}
Compose: \quad (\mathbf{I}, \mathbf{S}^0) \quad &\mapsto \quad x^g, \\
s.t. \quad x^g \quad &\in \quad \mathbf{I}, \\
|\mathbf{S}^0| \quad &= \quad 3, \\
|\mathbf{I}| \quad &= \quad 5
\end{aligned}
\tag{7}
$$

*Composition Evaluation Dataset*

A set of tuple $\mathcal{Q} = \{(\mathbf{S}_i^0, \mathbf{I}_i, \mathbf{Y}_i)\}$ is needed to evaluate the constrained composition algorithm in Eqn. (7). In the tuples $\mathcal{Q}$, the label set $\mathbf{Y}_i \subset \mathbf{I}_i$ denotes the positive items for the query $(\mathbf{S}_i^0, \mathbf{I}_i)$, *i.e.* the output $x_i^g$ is a good hit if it is in the set $\mathbf{Y}_i$.

Given the evaluation set $\mathcal{Q}$, we use the following accuracy measurement to evaluate the performance of the different models,

$$
accuracy = \frac{1}{|\mathcal{Q}|} \sum_{i \in \mathcal{Q}} I(x_i^g \in \mathbf{Y}_i),
\tag{8}
$$

where $I(*)$ is the indicator function.

The set of evaluation tuples can be built from the test dataset $(\mathbf{S}_i, y_i)$ with the following formula,

$$
\mathcal{Q}^{\text{auto}} = \{(\mathbf{S}_i - \mathbf{S}_i^1, \mathbf{I}_i^4 \cup \mathbf{S}_i^1, \mathbf{S}_i^1 | y_i = 1\},
\tag{9}
$$

where $\mathbf{I}$ denotes all fashion items in the database, $\mathbf{I}_i^4$ means a 4-item set randomly sampled from $\mathbf{I}$, and $\mathbf{S}_i^1$ denotes a 1-item set randomly sampled from $\mathbf{S}_i$. In other words, for each "good" outfit, randomly sample one item as the target item and sample four items from the database as the confusion items.

The evaluation set $\mathcal{Q}^{\text{auto}}$ built by Eqn. (9) treats the actual item $\mathbf{S}_i^1$ as the only positive item, and all the randomly sampled items $\mathbf{I}_i^4$ as negatives. However, this may not be the case, because randomly sampled items can potentially go very well with the seed outfit $S^0$. Therefore, we employ our fashion experts to label a golden standard evaluation set $Q^{gold}$:

$$Q^{gold} = \{(\mathbf{S}_i - \mathbf{S}_i^1, \mathbf{I}_i^4 \cup \mathbf{S}_i^1, \mathbf{S}_i^g | y_i = 1\}, \quad (10)$$

where the positive items are replaced by the golden standard $\mathbf{S}_i^g$ provided by fashion experts.

| $Q^{auto}$ | full | image | title | categories |
|---|---|---|---|---|
| $Q^{auto}$ | **0.50** | 0.42 | **0.50** | 0.35 |
| $Q^{gold}$ | 0.72 | 0.69 | 0.69 | **0.77** |

TABLE IX: Accuracy of the composition model. The accuracy is defined in Eqn. (8), the evaluation set $Q^{auto}$ is defined in Eqn. (9), and the golden standard evaluation set is defined in Eqn. (10)

The number of golden standard evaluation tuples labeled by the experts is 150. A closer look into the golden standard evaluation set $Q^{gold}$ reveals that,

1) On average, two items are labeled by the fashion experts as positive for each evaluation set, *i.e.* avg($|\mathbf{S}_i^g|$) = 2.
2) 93.7% of the actual item $\mathbf{S}_i^1$ (as defined in Eqn. (9)) is labeled as positive by the experts, which validates the use of user engagement levels as the outfit quality indicator.

While asking the fashion experts to label the matches, we also ask them to explain the decisions. There are certain factors frequently mentioned by the experts, for example, *color tone*, *color contrast*, *pattern prints*, *categorical matches*, and *coherence*. Ideally, all these factors can be captured by the image alone, but due to the limitation of the current image understanding capabilities and limited data samples, the meta-data is more powerful for recognizing certain fashion factors, such as the categorical matches and pattern prints.

### *Fashion Outfit Composition Results*

Table IX shows the accuracy comparison of the composition models using different item modalities, and Table X shows some examples of the constrained composition results. The result shows that,

1) The best composition model can simulate the expert results with 77% accuracy, which shows great potential of this work. However, as indicated by the constrained process in Eqn. (7) and the limited evaluation sample size, the evaluation is very constrained, and future work is needed to make the human and machine comparison more fair. The ideal evaluation is to build an online system, and let the users to rate the auto generated outfits, but that is out of the scope of this paper, and we plan it to the future work.
2) As with the scorer model evaluation, using both the images and the meta-data achieves the best performance (*full*) for both evaluation sets.
3) The performance on the original evaluation set $Q^{auto}$ correlates well with that of the golden standard evaluation set, which gives good justification to use the larger $Q^{auto}$ in future work.

### *D. Model Analysis*

In this section, we present further analysis of the core outfit scoring component. We first analyze how the scoring model attend to each item in the fashion outfit. Then, we present some error analysis for the scoring model.

TABLE X: Automatic fashion outfit composition results. Each row include the seed items $\mathbf{S}_i^0$, and the sorted candidate items $\mathbf{I}_i$. The candidate sets in the four rows are sorted by the modality configurations, full, image, title, categories, respectively. "full" means combining all three modalities. Using the scorer model, the items that goes best with the seed items are listed first. The highlighted candidate item is the pivot item $S_i^1$ in Eqn. (9), *i.e.* the ground truth item of the evaluation set $Q^{auto}$. Therefore, higher ranked highlighted item means better performance.

### *Fashion Item Importance*

A fashion outfit consists of multiple fashion items, and these items contribute differently to the fashion outfit quality. Understanding these differences can be very helpful, because it gives the model some interpretabilities, and more importantly, help the users to curate better outfits.

For the fashion outfits, we can analyze the item importance assigned by a scoring model with the Algorithm 2. The algorithm basically computes the quality score decrements after replacing the item with an arbitrary item in the database, and larger decrements indicates higher importance.

Table VIII shows the examples of the importance orders predicted by different modality configurations. These examples show that different modalities pick up different cues to predict outfit quality. From the examples, we make the following observations,

1) It is safe to replace the least important item for all the three examples, in order to improve outfit quality. Take the third outfit as an example, the item with the model appears to be the least important item in the outfit, because it overlaps with the other items.
2) The *"image"* model tends to pick up the color tone and shape. For example, the white dress of the first example does not match well well with the other items in terms of colors.
3) The *"image"* model tends to agree with the *"full"* model on the importance order, which means the appearance

---

**Algorithm 2:** Fashion item importance ordering

---

**Input** : Outfit scoring model $f(\mathbf{S}_i; \theta)$
**Input** : A fashion outfit $\mathbf{S}_i \triangleq \{x_{i1}, x_{i2}, \ldots, x_{i|\mathbf{S}_i|}\}$
**Output** : The item list $[x_{i,j^1}, x_{i,j^2}, \ldots, x_{i,j^{|S_i|}}]$ ordered by importance

**1** Initial score $p_i^0 \leftarrow f(\mathbf{S}_i; \theta)$
**2 foreach** *Fashion item* $x_{i,j} \in \mathbf{S}_i$ **do**
**3**   Random sample one item from the database $x_{i,j}^r \in \mathbf{I}$
**4**   Replace $x_{i,j}$ from $\mathbf{S}_i$ with $x_{i,j}^r$:
     $\mathbf{S}_i^j \leftarrow \mathbf{S}_i - \{x_{i,j}\} + \{x_{i,j}^r\}$
**5**   Score of the new outfit: $p_i^j \leftarrow f(\mathbf{S}_i^j; \theta)$
**6**   Outfit quality decrements: $d_i^j \leftarrow p_i^0 - p_i^j$
**7 end**
**8 return** *sort* $\mathbf{S}_i$ *by* $\{d_i^j | j = 1 \ldots |\mathbf{S}_i|\}$

---

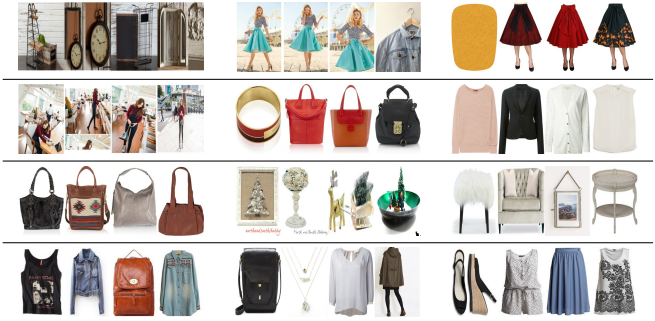

TABLE XI: Error examples of the top 1000 predictions. The four rows correspond to the full, image, title, categories modality configurations, respectively. The error rates are 0.353, 0.415, 0.371, 0.605, respectively.

determines the overall style of the outfit.

4) It it hard to find consistent reasons of why items are ranked higher for the *"title"* and *"category"* models.
5) The orders of the *"full"* model can not be explained by combinations of the orders of the other single modality models, which means the interaction between the modalities is non linear and complex.

It can be useful for users by suggesting to replace the least significant items. For example, for the second set of Table VIII, the third item can be replaced because of its lowest importance score predicted by the *full* and *title* model, which makes sense because its color is not compatible with other items. We plan to make further quantitative analysis to support these claims in follow up works.

### *Error case analysis*

It is important for future works to understand the error cases of the scoring models. We order the test fashion outfits by the predicted quality scores. We investigate the top 1000 fashion outfits that actually do not accept user engagement. The error examples and the error rates of various modality configurations are shown in Table XI, which shows that the trend of the error rates are comparable with the AUC and AP results in Table IV, *e.g.*, the full modality configuration produces the least errors in the top 1000 list. The errors made by the *image* model tend to be more visually consistent, and the errors made by the *title* and *category* tend to be more consistent in terms of the item semantics.

### *E. Implementation Details*

For the optimization, we use Adam optimizer with a batch size of 50, learning rate of 0.01, and 40K iterations, and we half the learning rate for every 15K iterations. For the *GloVe* model used in the item title encoder, we use the model pretrained on webpage corpus with 42 billion tokens and 1.9 million unique words, and we fix it during the training. The entire framework is implemented with TensorFlow [32].

## V. CONCLUSIONS

In this paper, we consider the challenging problem of fashion outfit composition, which reflects the difficulties of matching domain expert knowledge and modeling the diversity in fashion. We propose a generic composition algorithm based on outfit quality scorer. The outfit quality scorer is an end-to-end trainable system, which achieves promising performance. We find that the combination of multi-modalities and proper pooling of the instance level features, leads to the best performance. In the future, we plan to collect more data and find better ways to evaluate the composition algorithm.

## VI. ACKNOWLEDGMENT

We thank Hu Hao and Eujin Rhee from Polyvore for their help in this work. This work was supported in part by New York State through the Goergen Institute for Data Science at the University of Rochester, and a FREP award from Yahoo Research.